\documentclass[aps,pra,10pt,twocolumn,nofootinbib,superscriptaddress]{revtex4-1}
\usepackage[latin1]{inputenc}
\usepackage[english]{babel}
\usepackage{graphicx}
\usepackage{color}
\usepackage{amsmath}
\usepackage{amssymb}
\usepackage{multirow}
\usepackage{color}

\begin{document}

\title{Modeling anisotropic magnetized White Dwarfs with $\gamma$ metric}

\author{D. Alvear Terrero\footnote{dianaalvear@icimaf.cu}}
\email{dianaalvear@icimaf.cu}
\affiliation{Instituto de Cibern\'{e}tica, Matem\'{a}tica y F\'{\i}sica, \\
 Calle E esq a 15, Vedado 10400, La Habana Cuba}

\author{V. Hern\'andez Mederos\footnote{vicky@icimaf.cu}}
\email{vicky@icimaf.cu}
\affiliation{Instituto de Cibern\'{e}tica, Matem\'{a}tica y F\'{\i}sica, \\
 Calle E esq a 15, Vedado 10400, La Habana Cuba}

\author{S. L\'opez P\'erez}
\email{slopez@estudiantes.fisica.cu}
\affiliation{Facultad de F{\'i}sica, Universidad de la Habana,\\ San L{\'a}zaro y L, Vedado, La Habana 10400, Cuba}

\author{D. Manreza Paret}
\email{dmanreza@fisica.uh.cu}
\affiliation{Facultad de F{\'i}sica, Universidad de la Habana,\\ San L{\'a}zaro y L, Vedado, La Habana 10400, Cuba}
\affiliation{
Instituto de Ciencias Nucleares, Universidad Nacional Aut\'onoma de M\'exico,\\ Apartado Postal 70-543, CdMx 04510, M\'exico}

\author{A. P\'erez Mart\'{\i}nez}
\email{aurora@icimaf.cu}
\affiliation{Instituto de Cibern\'{e}tica, Matem\'{a}tica y F\'{\i}sica, \\
	Calle E esq a 15, Vedado 10400, La Habana Cuba}

\author{G. Quintero Angulo}
\email{gquintero@fisica.uh.cu}
\affiliation{Facultad de F{\'i}sica, Universidad de la Habana,\\ San L{\'a}zaro y L, Vedado, La Habana 10400, Cuba}

\begin{abstract}
 The effect of magnetic fields in the Equations of State (EoS) of compact objects is the splitting of the pressure in two components, one parallel and the other perpendicular to the magnetic field. This anisotropy suggests the necessity of using structure equations considering the axial symmetry of the magnetized system. In this work, we consider an axially symmetric metric in spherical coordinates, the $\gamma$-metric, and construct a system of equations to describe the structure of spheroidal compact objects. In addition, we connect the geometrical parameter $\gamma$ linked to the spheroid's radii, with the source of the anisotropy. So, the model relates the shape of the compact object to the physics that determines the properties of the composing matter. To illustrate how our structure equations work, we obtain the mass-radii solutions for magnetized White Dwarfs.
 Our results show that the main effect of the magnetic field anisotropy in White Dwarfs structure is to cause a  deformation of these objects. Since this effect is only relevant at low densities, it does not affect the maximum values of magnetized White Dwarf's masses, which remain under Chandrasekhar limit.
\end{abstract}

\maketitle


\section{Introduction}\label{intro}

Magnetic fields are present in almost all stars during their stellar evolution, becoming huge in the final stage, when they turn into compact objects.
Measurements of  periods and spin down of soft-gamma repeaters (SGR) and X-ray luminosities of anomalous X-ray pulsars (AXP) \cite{Woods2006csxs}, support the idea of the existence of magnetars: neutrons stars with surface magnetic fields as large as $10^{14}-10^ {16}$~G \cite{Duncan:1992hi}. In the case of White Dwarfs (WDs), observed surface magnetic fields range from $10^6~$G to $10^9~$G \cite{2015SSRv..191..111F}. Although the inner magnetic fields can not be observed  directly, their bounds can be estimated with theoretical models based on macroscopic and microscopic analysis. The  maximum magnetic fields are around $10^{13}~$G for WDs \cite{1674-4527-15-10-1735,Diana} and about $5\times 10^{18}~$G for neutron stars \cite{1674-4527-15-7-975}.

From a microscopic point of view, a magnetic field acting on a fermion gas breaks the spherical symmetry and produces an anisotropy in the quantum-statistical average of the energy-momentum tensor. The effect of this anisotropy is the splitting of the pressure into two components, one along the magnetic field ---the parallel pressure  $P_\parallel$--- and another in the transverse direction ---the perpendicular pressure $P_\perp$---, so that $T^{\mu}_{\nu} = {\rm diag}(E, -P_{\perp}, -P_{\perp}, -P_{\parallel})$. Consequently, a gas of fermions under the action of a constant and uniform magnetic field has an anisotropic ---axially symmetric--- equation of state (EoS) \cite{2000PhRvL..84.5261C}. For this reason, when modeling the structure of magnetized compact objects, one should consider axial symmetry instead of the spherical symmetry used when solving the Tolman-Oppenheimer-Volkoff (TOV) equations.

Our first attempt addressing this issue, on Refs.~\cite{1674-4527-15-10-1735,1674-4527-15-7-975,Diana}, was to consider a metric in cylindrical coordinates $(t,r,\phi,z)$  to obtain Einstein's field equations following the procedure described in Ref.~\cite{trendafilova}. This model lead us to obtain some information about the effects of the magnetic field in terms of the shape  ---prolateness or oblateness--- of the compact object as well as upper limits for the values of the magnetic field that can sustain these stars ($B_u \simeq 10^{13}~$G for WDs \cite{1674-4527-15-10-1735}). However, since we assumed that all the magnitudes depend only on the radial coordinate $r$, we were unable to determine the total mass.

Therefore, we return to spherical coordinates. Let us remark that anisotropies in the energy-momentum tensor are admitted in spherical symmetry as long as the tensor has the form  $T^{\mu}_{\nu} = {\rm diag}(E, -p_{r}, -p_{t}, -p_{t})$, where $p_{r}$ is a radial pressure and  $p_{t}$ is a tangential one \cite{Dev:2000gt,Harko:2002db}. However, this is not compatible with the anisotropy due to magnetic fields. Thus, we are going to use an axially symmetric metric in  spherical coordinates to account for the magnetic anisotropy of the system.

Hence, in this work, we start from a metric with a $\gamma$ parameter associated to the deformation of the stars. This metric was previously presented in \cite{Zubairi:2017yna, Zubairi:2017gvp} and allows to obtain a set of structure equations that generalize the TOV equations to axially symmetric objects. The novelty of our treatment consists in computing the total mass as for a spheroidal object and proposing an ansatz to relate $\gamma$ with the ratio between the central  pressures, which connects the physics of the system with its geometry.

As an example and test case, we solve these anisotropic structure equations for magnetized WDs for different values of the magnetic field: $10^{12}~$G, $10^{13}~$G and $10^{14}~$G, which cover both the weak and the strong magnetic field regimes. Motivated by the interest they have rised as potential sources of super-Chandrasekhar WDs \cite{Das:2013gd}, we tackle strongly magnetized WDs ($B \gtrsim B_u$). Our results support that weakly magnetized WDs are more realistic, which reinforces the existence of the previously obtained threshold $B_u$.

In Section \ref{sec2} we present magnetized WDs EoS and discuss the magnetic field effects on the energy density and pressure. Section \ref{sec3} is devoted to TOV solutions while the anisotropic structure equations are presented in Section~\ref{sec4}. Corresponding numerical results for magnetized WDs and their discussion can be found in Section~\ref{sec5} and concluding remarks in Section~\ref{sec6}.

\section{EoS for magnetized White Dwarfs}\label{sec2}

Typical WDs are composed by carbon or oxygen atoms. The role of the different particles conforming these atoms in the star's physics depends on their masses. Due to its relative low mass, only the degenerated gas of relativistic electrons determine the pressure that compensates the gravitational collapse of the star. The heavier neutrons and protons behave non-relativistically, and contribute mainly to the mass and energy density.

The pressures and the energy density of the electron gas in magnetized WDs are obtained starting from the thermodynamical potential \cite{2000PhRvL..84.5261C} \footnote{All expressions in this Section are in natural units, where $\hbar=c=1$.}:
\begin{multline}\label{Thermo-Potential}
\Omega(B,\mu,T)=-\frac{e B}{2\pi^2}\int_{0}^\infty dp_3\sum_{l=0}^{\infty}g_l \bigg[ \varepsilon_l \\
 +T\ln \left(1+e^{-(\varepsilon_l-\mu)/T}\right)\left(1+e^{-(\varepsilon_l+\mu)/T}\right)\bigg],
\end{multline}
being $\varepsilon_l=\sqrt{p_3^2+2|eB|l+m^2}$ the electron spectrum in a magnetic field. In Eq.~(\ref{Thermo-Potential}) the magnetic field ${\bf B}$  is supposed uniform, constant and in the $z$ direction, $l$ stands for the Landau levels and the factor $g_l=2-\delta_{l0}$ includes the spin degeneracy of the fermions for $l\neq0$. $T$ is the  absolute temperature, $\mu$ the chemical potential, $m$ is the electron mass and $e$ its charge.

Note that, in general, the thermodynamical potential on Eq.~(\ref{Thermo-Potential}) can be divided in two contributions
\begin{equation}\label{Thermo-Potential-2}
\Omega (B,\mu,T)=\Omega^\text{vac}(B)+\Omega^\text{st}(B,\mu,T).
\end{equation}

The second term, $\Omega^\text{st}(B,\mu,T)$, arises from statistical considerations and reads
\begin{multline}\label{Thermo-Potential-st}
\Omega^\text{st}(B,\mu,T)=-\frac{e B}{2\pi^2}\int_{0}^\infty dp_3\sum_{l=0}^{\infty}g_l \\ \times \bigg[T\ln \left(1+e^{-(\varepsilon_l-\mu)/T}\right)\left(1+e^{-(\varepsilon_l+\mu)/T}\right)\bigg].
\end{multline}

When studying WDs, since the surface temperatures detected are much smaller than the Fermi temperature, it is accepted to consider the degenerate limit for the fermion gas ($T\to 0$) to compute the thermodynamical potential \cite{shapiro,camenzind}. In that case, the statistical term becomes
\begin{equation}\label{Thermo-Potential-st1}
  \Omega^\text{st}(B,\mu,0) = -\frac{eB}{2\pi^2}\int_{0}^\infty \!\!\!\!dp_3\sum_{l=0}^{\infty}g_l(\mu-\varepsilon_l)\Theta(\mu-\varepsilon_l),
\end{equation}
where $\Theta(\zeta)$ is the unit step function. From the expression (\ref{Thermo-Potential-st1}), we obtain
\begin{equation}\label{Thermo-Potential-Sest}
\Omega^\text{st} (B,\mu,0)= \frac{m^2}{4\pi^2}\frac{B}{B_c}\!\sum_{l=0}^{l_{max}}g_{l}\!\left[ \mu\,p_F -\varepsilon_l^2\ln\!\left(\frac{\mu+ p_F}{\varepsilon_l}\right)\right]\!,
\end{equation}
where $l_{max}= I[\frac{\mu^2-m^2}{2eB}]$, $I[z]$ denotes the integer part of $z$ and the Fermi momentum is ${p_F}=\sqrt{{\mu}^2-\varepsilon_l^2}$.

On the other hand, the term $\Omega^\text{vac}(B)$ in Eq.~(\ref{Thermo-Potential-2}):
\begin{equation}\label{Thermo-Potential-vac}
\Omega^\text{vac}(B)=-\frac{e B}{2\pi^2}\int_{0}^\infty dp_3\sum_{l=0}^{\infty}g_l \varepsilon_l,
\end{equation}
does not depend on the chemical potential nor on the temperature and corresponds to the vacuum. This contribution presents an ultraviolet divergence that must be renormalized \cite{1951PhRv...82..664S}. Depending on the  value of $B$ with respect to the critical magnetic field for electrons, $B_c=m^2/e = 4.4\times10^{13}$~G (Schwinger field)\footnote{The magnetic field at which the cyclotron energy of the electrons is comparable to its rest mass.}, the renormalization of $\Omega^\text{vac}$ leads to one of the following expressions corresponding to the weak $(B<B_c)$ and strong magnetic field ($B>B_c$) limits \cite{2015PhRvD..91h5041F}
\begin{subequations}\label{Thermo-Potential-vac1}
\begin{eqnarray}
\Omega^\text{vac}_w (B,0,0) &= -\frac{m^4}{90 (2\pi)^2}  \left(\frac{B}{B_c}\right)^{\!4}\!, & \quad B<B_c \\ \label{Thermo-Potential-Svac}
\Omega^\text{vac}_s (B,0,0) &= \frac{(eB)^2}{24\pi^2(\hbar c)^3}\ln\frac{eB}{m^2}, & \quad B>B_c.
\end{eqnarray}
\end{subequations}

 As pointed out previously, the maximum magnetic field estimated for WDs interiors is around $10^{13}~$G, a value that is of the same order of $B_c$. In astrophysical scenarios the energy scales are determined by the temperature and the density. Then, for WDs in the zero temperature limit, the parameter that will set the relative relevance of magnetic field effects on the system is the density. If we consider typical values of densities for WDs and magnetic fields in weak regime, the system is characterized by the relation $eB \ll m^2 \ll \mu^2$. In this case, the vacuum contribution in Eq.~(\ref{Thermo-Potential-vac1}) can be neglected when compared to the statistical one in Eq.~(\ref{Thermo-Potential-Sest}). Therefore, the thermodynamical potential of the electron degenerate system can be approximated to $\Omega(B,\mu,0) = \Omega^\text{st}(B,\mu,0)$ when working below the Schwinger magnetic field.

In this regime, the distance between Landau levels \mbox{$(\sim eB)$} is small and we can consider the discrete spectrum as a continuum. This allows us to replace the sum over $l$ in Eq.~(\ref{Thermo-Potential-Sest}) by an integral through the Euler-MacLaurin formula \cite{1972hmfw.book.....A}
\begin{multline}
	\label{eq:EMcL}
  \frac{eB}{2} \sum_{l=0}^\infty g_l f(2eBl) \approx eB\!  \int_0^\infty \!\!\! f(2eBl)dl+ \frac{eB}{2}f(\infty) \\
	+ \sum_{k=1}^\infty \frac{2^{2k-1} }{(2k)!} (eB)^{2k} B_{2k} \left[f^{2k-1}(\infty)-f^{2k-1}(0)\right],
\end{multline}
where $f(2eBl) = (\mu-\varepsilon_l)\Theta(\mu-\varepsilon_l)$ and the coefficients $B_n$ stand for the Bernoulli numbers ($B_2=1/6$). Then, we can expand Eq.~(\ref{Thermo-Potential-st1}) onto the second power on $eB$, and take the classical limit by means of the change of variables $p_\perp^2= 2eBl=p_x^2+p_y^2$, with $p_\perp dp_\perp =eBdl$ \cite{1968PhRv..173.1210C}. Hence, we get the statistical part of the thermodynamical potential as follows
\begin{multline}\label{Thermo-Potential-st2}
	\Omega_{st}(B,\mu,0) = - \frac{m^4}{12\pi^2} \left[\frac{\mu \sqrt{\mu^2-m^2}}{m^2}\left(\frac{\mu^2}{m^2}-\frac{5}{2}\right)\right.
	\\
	+ \frac{3}{2} \ln\left(\frac{\mu+\sqrt{\mu^2-m^2}}{m}\right)
	\\
	\left.
	+ \left( \frac{B}{B_c} \right)^{2} \ln \left(\frac{\mu+\sqrt{\mu^2-m^2}}{m} \right) \right].
\end{multline}
From Eq.~(\ref{Thermo-Potential-st2}) we note that, in the weak magnetic field limit, the statistical part of the thermodynamical potential is expressed as a sum of two non-magnetic terms at $\mu \neq 0$ and $T=0$, plus a third term that depends also on the magnetic field.

Matter inside compact stars must be in stellar equilibrium. So, we must impose charge neutrality and baryon number conservation to the energy density and the pressures. With these considerations, the magnetized WDs EoS ---the pressures as a parametric function of the energy density--- becomes
\begin{subequations}\label{EoSEB}
\begin{align}
  E &= \Omega + \mu N +m_N\frac{A}{Z}N  +\frac{B^2}{8\pi},\label{EoS1}\\
  P_{\parallel} &= -\Omega-\frac{B^2}{8\pi}, \label{EoS2}\\
  P_{\perp} &= -\Omega-B\mathcal{M}+\frac{B^2}{8\pi},\label{EoS3}
\end{align}
\end{subequations}
where $N=-\partial\Omega/\partial\mu$ is the electron particle density and $\mathcal{M}=-\partial\Omega/\partial B$ the magnetization. Here, the thermodynamical potential $\Omega$ is given by Eq.~(\ref{Thermo-Potential-2}), with $\Omega^\text{vac}$  as in Eq.~(\ref{Thermo-Potential-vac1}) and $\Omega^\text{st}$ from Eq.~(\ref{Thermo-Potential-Sest}) or Eq.~(\ref{Thermo-Potential-st2}) according to the value of B. The term $ N m_N A/Z$ included in Eq.~(\ref{EoS1}) considers the contribution of the nucleons to the energy density\footnote{$m_N = 931.494$ MeV $\sim m_{n,p}$ and $A/Z$ is the number of nucleons per electron ($A/Z=2$ for carbon/oxygen WDs)}.

The last term in Eqs.~(\ref{EoSEB}) is the Maxwell contribution to the pressures and energy density, $P_{\perp}^B=E^B=-P_{\parallel}^B=B^2/8\pi$. Contrary to the case of Neutron Stars and Quark Stars \cite{Lattimer}, where the energy density and pressures are of the same order, WDs energy densities are three orders higher than the pressures. Therefore, the value delimiting when the Maxwell term becomes relevant is determined by the pressure, with $P_{\perp}^B$ comparable to the statistical pressures for magnetic fields higher than $1.78\times10^{11}$~G.\footnote{The magnetic energy density becomes relevant at $10^{14}~$G \cite{Terrero:2017mxc}.}

\begin{figure}[!ht]
	\includegraphics[width=0.48\textwidth]{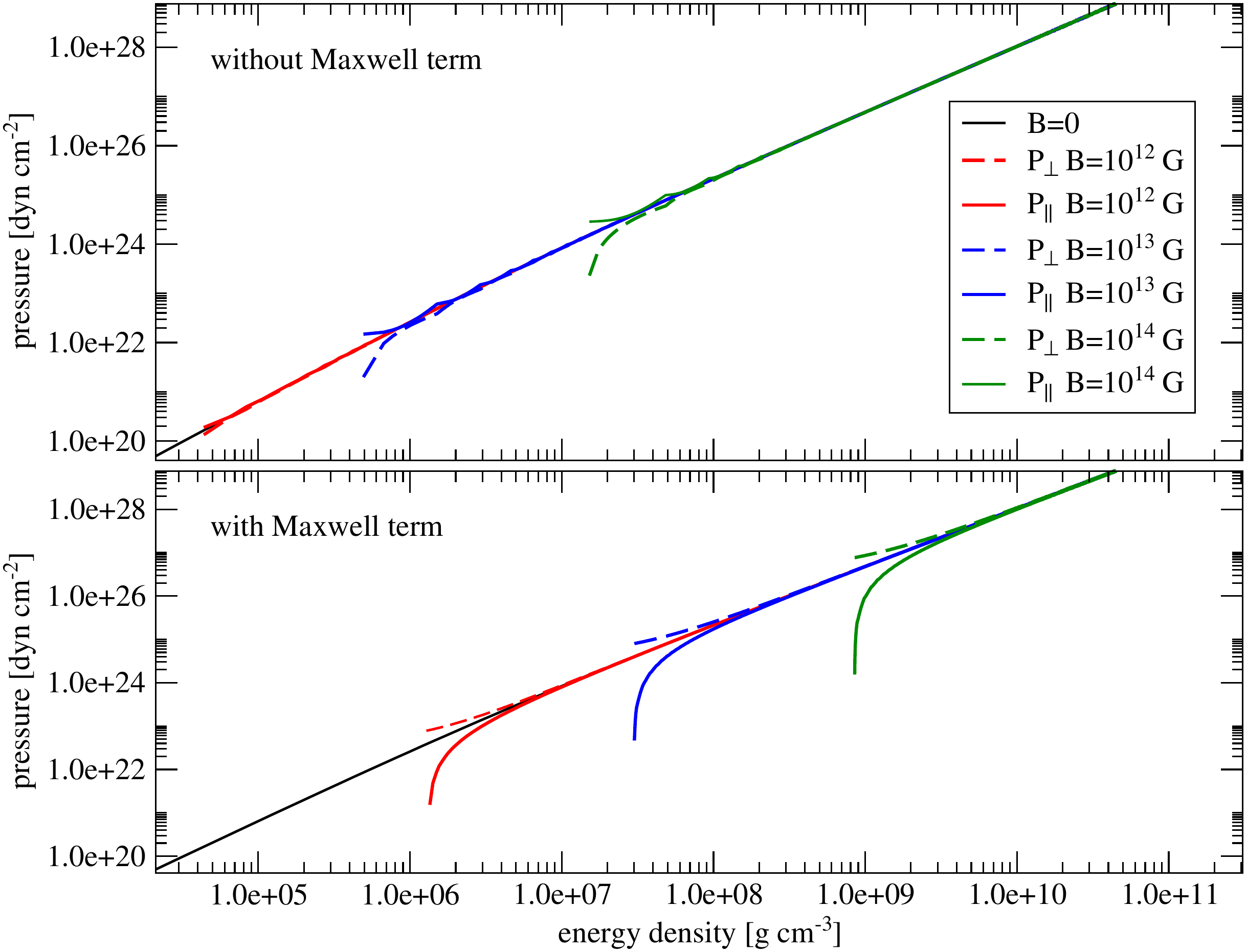}
	\caption{EoS for magnetized WDs at fixed values of the magnetic field for  $B=0$, $B=10^{12}$~G, $B=10^{13}$~G and $B=10^{14}$~G (in CGS units). In the upper panel Maxwell term is neglected while in the lower panel it is considered.}	\label{fig:EOS}
\end{figure}

Anyhow, we explore the parametric EoS given in Eqs.~(\ref{EoSEB}) and tackle two cases, the first one neglecting the Maxwell contribution (upper panel of Fig.~\ref{fig:EOS}) and the second one considering it (lower panel), for magnetized WDs with a carbon/oxygen composition at $B=0$, $B=10^{12}$~G, $B=10^{13}$~G and $B=10^{14}$~G. Note that at higher densities there is no appreciable difference between the perpendicular and the parallel pressures while at low densities the anisotropy starts to be noticeable, being the perpendicular pressure curve softer (harder) than the parallel one in the case without (with) Maxwell term.

\section{Magnetized WDs TOV solutions}\label{sec3}

\begin{figure}[ht]
	\centering
	\includegraphics[width=0.48\textwidth]{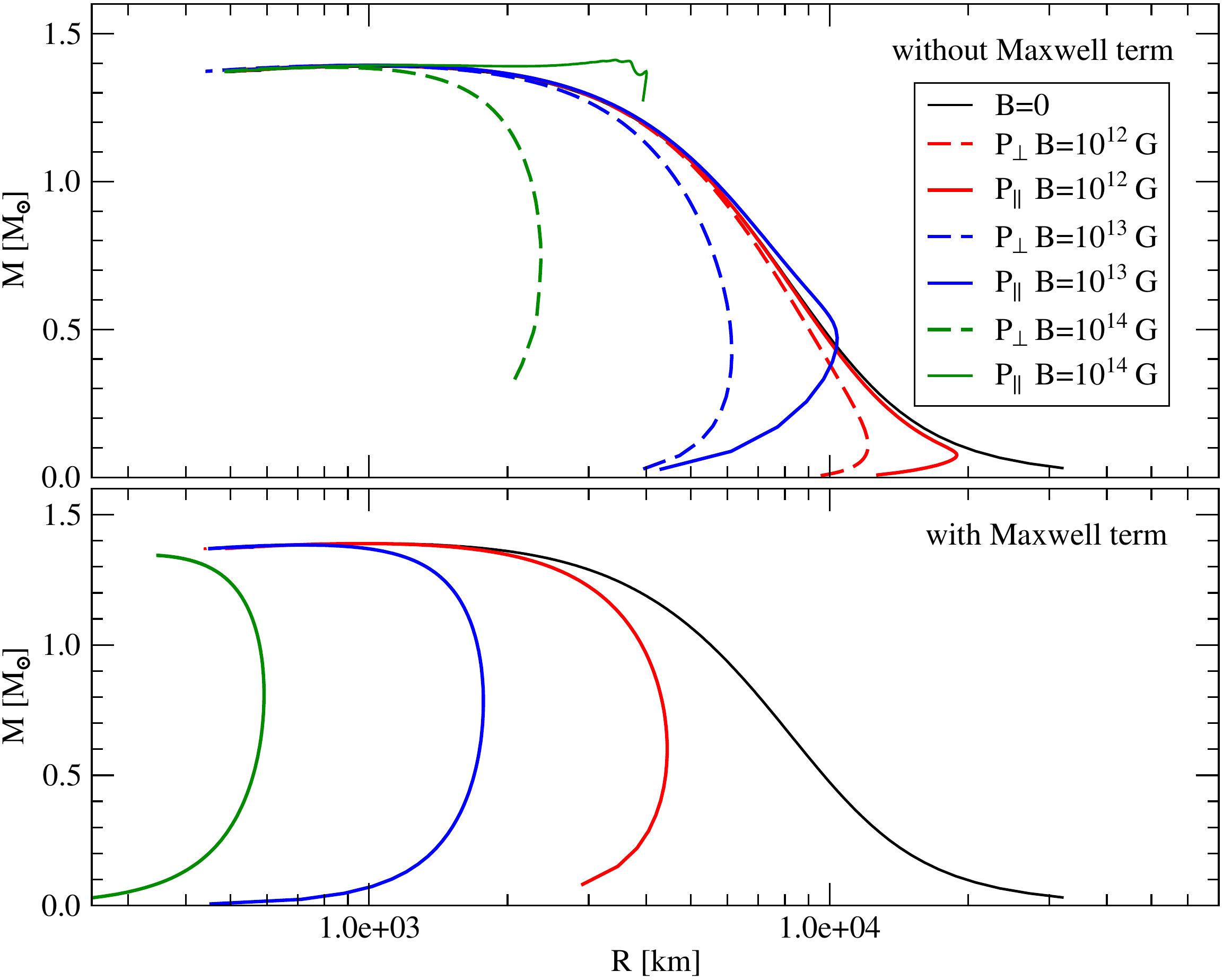}
\caption{Isotropic TOV equations solutions for the perpendicular and parallel pressures independently at $10^{12}~$G, $10^{13}~$G and $10^{14}~$G compared to the isotropic $B=0$ curve. The anisotropy becomes important in the low density regime.}
\label{FigIsotropicTOV}
\end{figure}

The magnetized WDs EoS obtained in previous section can be used to solve the standard isotropic TOV equations. In Fig.~\ref{FigIsotropicTOV}, we present the mass-radius curves obtained considering the pairs $(E,P_\parallel)$ and $(E, P_\perp)$ as independent EoS, as well as the corresponding non-magnetized curve \cite{1674-4527-15-10-1735}.

Despite the small differences between the pressures, the fact of using one or the other leads to different mass-radius relations for less dense systems.  Fig.~\ref{FigIsotropicTOV} shows that softer EoS produces smaller radius, which means the TOV solutions with lower values of pressures yield lower values of radii, as it is the case for the perpendicular (parallel) pressure when neglecting (considering) the Maxwell contribution. In the latter case (bottom panel of Fig.~\ref{FigIsotropicTOV}) perpendicular pressures solutions exist only for high densities and cannot be distinguished from the corresponding parallel pressures curves. These results highlight the importance of a model properly considering the anisotropy in the system.

In the first attempt to address the anisotropic structure, we used cylindrical symmetry and  obtained also that the pressures parallel and perpendicular inside the stars go to zero for different values of equatorial radius, where lower central pressures give lower radii \cite{1674-4527-15-10-1735}.

\section{$\gamma$-Metric and structure equations for magnetized compact objects}\label{sec4}

We devote this section to construct a general model suitable to study the structure of axially deformed compact objects. Our model is based on Refs.~\cite{Herrera:1998eq, Zubairi:2017yna, Zubairi:2017gvp}, where the authors show that a deformed compact object with axial symmetry can be described by  the metric
\begin{align}\label{Ec(13)}
	ds^2 = & - \left[1-2M(r)/r\right]^{\gamma}dt^2 + \left[1-2M(r)/r\right]^{-\gamma}dr^2 \nonumber \\
                 &+ r^2\sin\theta d\phi^2 + r^{2}d\theta^2,
\end{align}
where $\gamma = z/r$ parametrizes the polar radius $z$ in terms of the equatorial one $r$. Considering the fact that using smaller pressures in TOV equations leads to smaller radii as well as our interest in the effects on the structure equations coming from the magnetic field and the related anisotropy, we propose to interpret $\gamma$ as the ratio between the parallel and perpendicular central pressures, $P_{\parallel_0}$ and $P_{\perp_0}$ respectively
\begin{equation}
\gamma=\frac{P_{\parallel_0}}{P_{\perp_0}}. \label{gamma}
\end{equation}
This assumption is a first attempt to consider the anisotropy of the magnetized gas properly and allows us to connect the geometry with the physics of the system, implying that the shape of the star is only determined by the anisotropy of the EoS in its center. So, we are neglecting the fact that the star's deformation (i.e. the difference between the polar and the equatorial radii) also depends on the inner profiles of the anisotropic pressures. The approximation yields reasonable results for small deformations, i.e. $\gamma$ close to 1, as can be seen in Section~\ref{sec5}  for typical densities and magnetic field values of WDs. However, a more advanced calculation should take into account the variation of the parameter $\gamma$ throughout the star, just as if considering a nested set of shells with constant value of $\gamma$.

Starting from this metric, the energy-momentum tensor of the magnetized gas and using the mass of a spheroid to compute the star's mass, we obtain the following structure equations
\begin{subequations}\label{gTOV}
\begin{eqnarray}  \label{gTOV1}
&& \frac{dM}{dr}=4 \pi r^{2}\frac{(E_{\parallel} +E_{\perp})}{2}\gamma,\\ \nonumber
&&\frac{dP_{\parallel}}{dz}=\frac{1}{\gamma}\frac{dP_{\parallel}}{dr}\\  \label{gTOV3}
&&\phantom{\frac{dP_{\parallel}}{dz}}=-\frac{(E_{\parallel}+P_{\parallel})[\frac{r}{2}+4 \pi r^{3}P_{\parallel}-\frac{r}{2}(1-\frac{2M}{r})^{\gamma}]}{\gamma r^{2}(1-\frac{2M}{r})^{\gamma}}, \qquad\quad\\ \label{gTOV2}
&&\frac{dP_{\perp}}{dr}=-\frac{(E_{\perp}+P_{\perp})[\frac{r}{2}+4 \pi r^{3}P_{\perp}-\frac{r}{2}(1-\frac{2M}{r})^{\gamma}]}{ r^{2}(1-\frac{2M}{r})^{\gamma}},
\end{eqnarray}
\end{subequations}
which describe the variation of the mass and the pressures with the spatial coordinates $r,z$ for an anisotropic axially symmetric compact object. Note that these equations are coupled through the dependence with the energy density and the mass.

Since the parallel pressure has its maximum central value at $z=0$ and goes to zero at the surface, we assume it depends just on the $z=\gamma r$ coordinate. The perpendicular pressure, on the contrary, is  zero at $r=0$, therefore depending on the radial coordinate.

In general terms, the solutions of Eqs.~(\ref{gTOV}) are computed similarly to how it is done usually for the TOV equations. In this case, we start from a point in the center with $E_0 = E(r=0)$, $P_{\parallel_0} = P_\parallel(r=0)$ and $P_{\perp_0} = P_\perp(r=0)$ taken from the EoS on Eq.~(\ref{EoSEB}). The equatorial and polar radii of the star, $R$ and $Z=\gamma R$, are respectively defined by $P_\parallel (Z) = 0$  and $P_\perp (R) = 0$, while the mass of the star is $M=M(R)$. In practice, this condition is established by the lower central pressure, which determines the value of the corresponding radius ($R$ if $P_{\perp_0}$ and $Z$ if $P_{\parallel_0}$), from where the other radius can be computed by means of $\gamma$.

There is also a remarkable difference with respect to the solution of standard TOV equations in the manner we compute the energy density from the EoS during the integration process. To clarify this point, let us denote as $c_1(\mu),c_2(\mu)$ the 2D parametric curves given by
\begin{subequations}
	\begin{eqnarray}
	c_1(\mu)&=&(E(\mu),P_{\parallel}(\mu)) \label{c1}\\
	c_2(\mu)&=&(E(\mu),P_{\perp}(\mu)) \label{c2}
	\end{eqnarray}
\end{subequations}
with $E(\mu),P_{\parallel}(\mu)$ and $P_{\perp}(\mu)$ defined by Eqs.~(\ref{EoSEB}). Given $\widetilde{P}_{\parallel}$ and $\widetilde{P}_{\perp}$, obtained in one integration step of Eqs.~(\ref{gTOV}), two parametric values $\widetilde{\mu}_{\parallel}$ and $\widetilde{\mu}_{\perp}$ are computed interpolating Eqs.~(\ref{EoS2}) and (\ref{EoS3}) respectively. The corresponding points in the curves (\ref{c1}) and (\ref{c2}) are $c_1(\widetilde{\mu}_{\parallel})=(\widetilde{E}_{\parallel},\widetilde{P}_{\parallel})$ and $c_2(\widetilde{\mu}_{\perp})=(\widetilde{E}_{\perp},\widetilde{P}_{\perp})$, where $\widetilde{E}_{\parallel}=E(\widetilde{\mu}_{\parallel})$ and $\widetilde{E}_{\perp}=E(\widetilde{\mu}_{\perp})$. Hence, in the next integration step, we update the right hand side of Eq.~(\ref{gTOV2}) using the point $c_1(\widetilde{\mu}_{\parallel})$ with $E=\widetilde{E}_{\parallel}$ and $P_{\parallel}=\widetilde{P}_{\parallel}$.  Similarly,  we update Eq.~(\ref{gTOV3}) with $c_2(\widetilde{\mu}_{\perp})$ by taking $E=\widetilde{E}_{\perp}$ and $P_{\perp}=\widetilde{P}_{\perp}$.

The use of different values of the energy density when integrating Eqs.~(\ref{gTOV2}) and (\ref{gTOV3}) is a consequence of canceling the dependence on the angular variables and assuming that $P_{\perp}$ evolves in the equatorial direction and $P_{\parallel}$ in the polar one and a warning about the fact that for a complete description of the anisotropic object one should consider a full tridimensional treatment.

The existence of two energies at each integration step introduces the puzzle of selecting which of them should be used to compute the total mass. Note that, since we are dealing with an anisotropic object the mass density is also anisotropic. Along the equatorial direction the mass density is equal to
\begin{eqnarray}\label{massdiff1}
	dM &=& 4 \pi \gamma r^2 E_{\parallel} dr,
\end{eqnarray}
while in the polar direction it reads
\begin{eqnarray}\label{massdiff2}
	dM &=& 4 \pi \frac{z^2}{\gamma^2} E_{\perp} dz.
\end{eqnarray}
In Eqs.~(\ref{massdiff1}) and (\ref{massdiff2}) we have used the parallel and the perpendicular energy density in regard of the differentiation direction. Now, taking into account that $z = \gamma r$, Eq.~(\ref{massdiff2}) can be transformed into
\begin{eqnarray}\label{massdiff3}
	dM &=& 4 \pi \gamma r^2 E_{\perp} dr.
\end{eqnarray}
Adding Eqs.~(\ref{massdiff1}) and (\ref{massdiff3}), we get
\begin{equation}\label{massdifffinal}
\frac{dM}{dr} = 4 \pi \gamma r^2 \frac{E_{\parallel} + E_{\perp}}{2}.
\end{equation}
Eq.~(\ref{massdifffinal}) indicates that, if we don't want to lose the information about the mass density anisotropy, we must update the right hand side of Eq.~(\ref{gTOV1}) with the average energy density $E=(\widetilde{E}_{\parallel}+\widetilde{E}_{\perp})/2$.
Note that small differences between $\widetilde{E}_{\parallel}$ and $\widetilde{E}_{\perp}$ during the integration process imply that the change between the pressures is not being drastically amplified so that the ratio $P_{\parallel}/P_{\perp}$ remains close to its value at the star's center and the ansatz is then justified.
This can be checked numerically by computing the variation with $r$ of the relative difference between the parallel and perpendicular energies. For all the most deformed cases we consider (solutions in Table~\ref{tab:data}), we obtained that $|\widetilde{E}_{\perp}-\widetilde{E}_{\parallel}|/E_0 \lesssim10^{-3}$.

The combination of the structure equations in Eqs.~(\ref{gTOV}) with the ansatz given by Eq.~(\ref{gamma}) allows us to describe the internal variations of the mass and the pressures of a magnetized compact object. It is important to remark that by setting $B=0$, the model automatically yields $P_\perp = P_\parallel$ and $\gamma=1$. This means that we recover the spherical TOV equations from  Eqs.~(\ref{gTOV}) and thus, the standard non-magnetized solution for the structure of compact objects.

In what follows, we study the solutions of Eqs.~(\ref{gTOV}) for magnetized WDs EoS, even though these structure equations describe any anisotropic axially deformed compact object provided that it is spheroidal.

\section{Magnetized WDs numerical results and important remarks}\label{sec5}

\begin{figure}[ht]
	\includegraphics[width=0.48\textwidth]{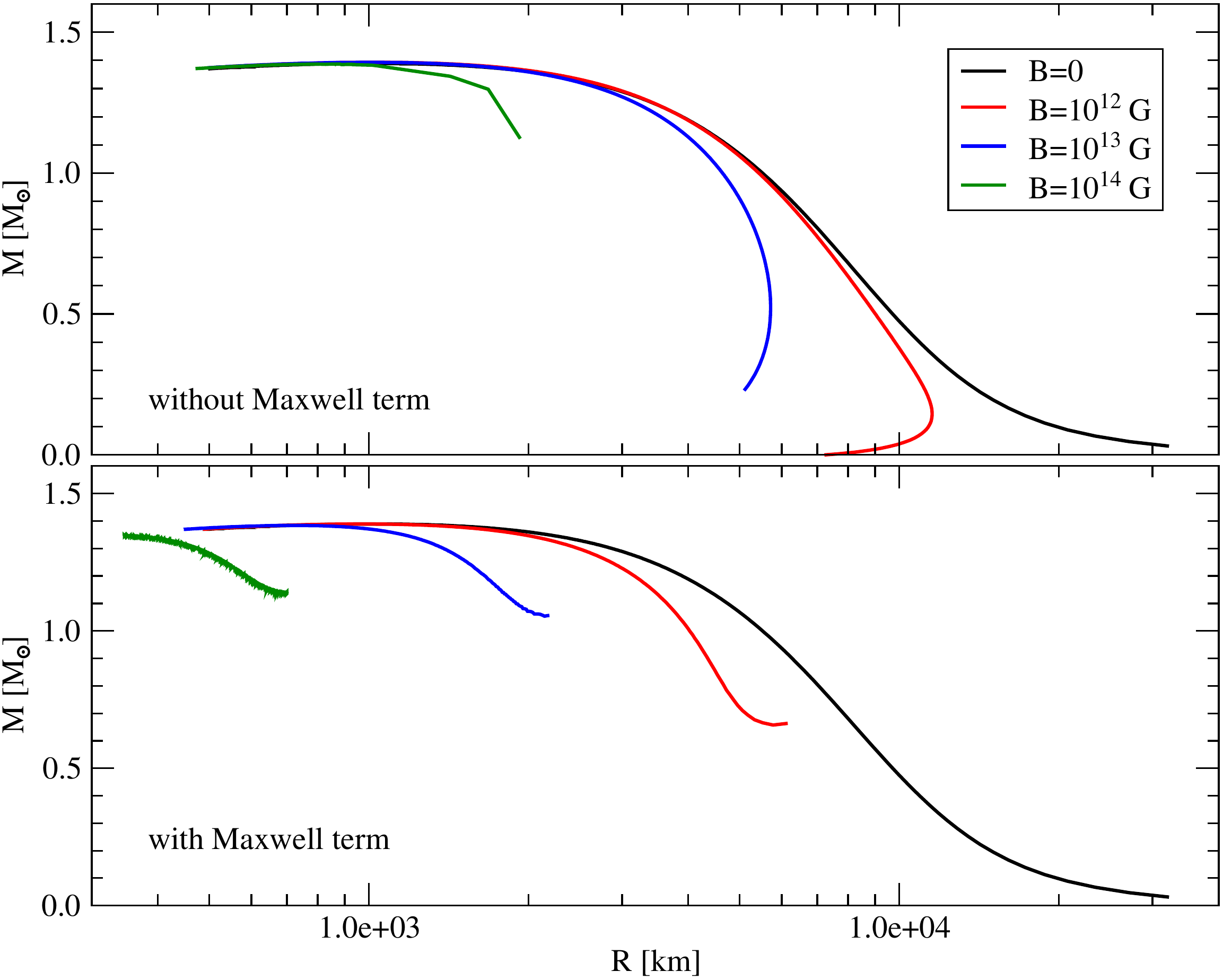}
	\caption{Mass versus the equatorial radius $R$. In the upper panel Maxwell term is neglected while in the lower panel it is considered.}
	\label{FigMR}
\end{figure}

\begin{figure}[ht]
	\includegraphics[width=0.48\textwidth]{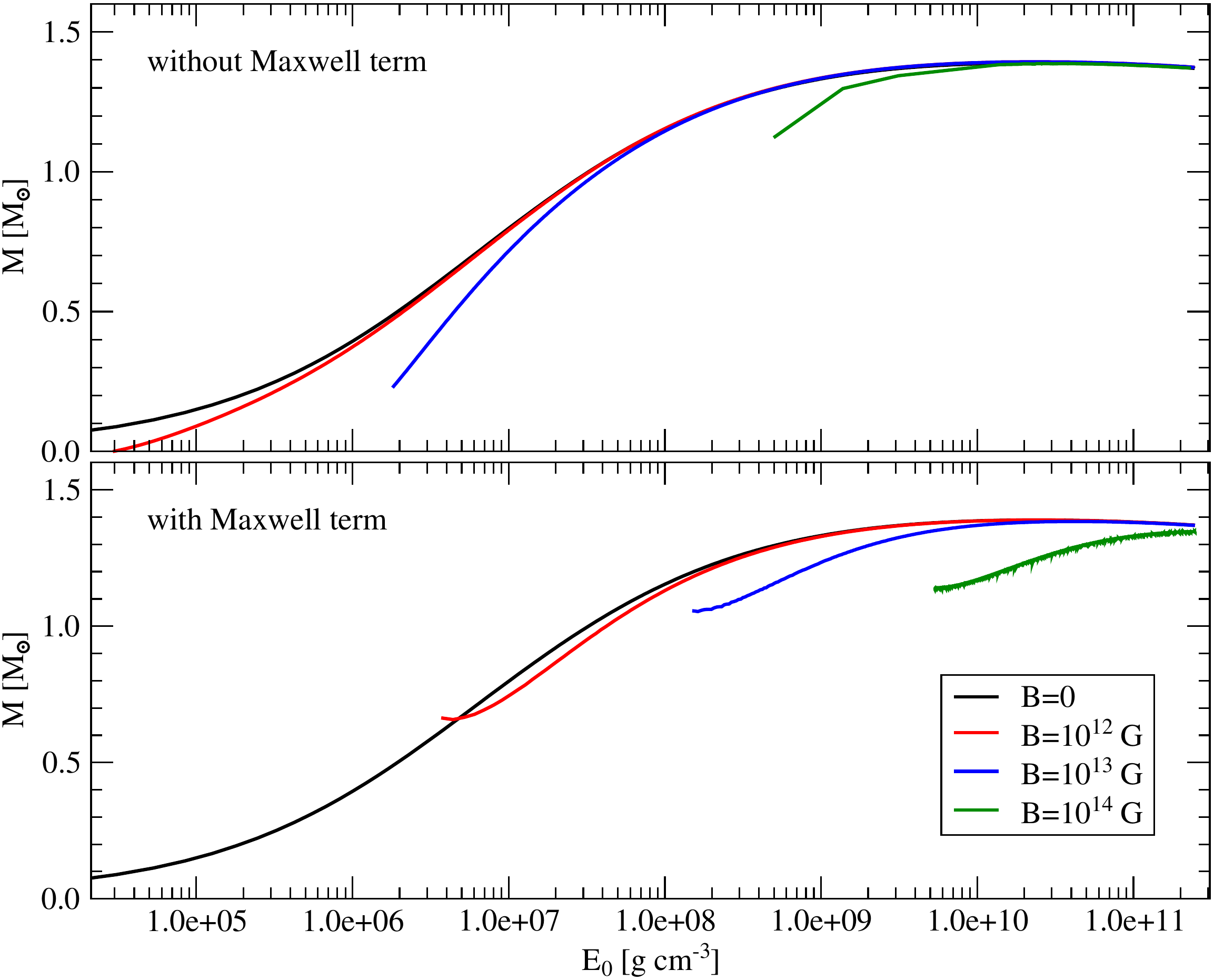}
	\caption{Mass as a function of central energy densities. In the upper panel Maxwell term is neglected while in the lower panel it is considered.}
	\label{FigME0}
\end{figure}

\begin{figure}[ht]
	\includegraphics[width=0.48\textwidth]{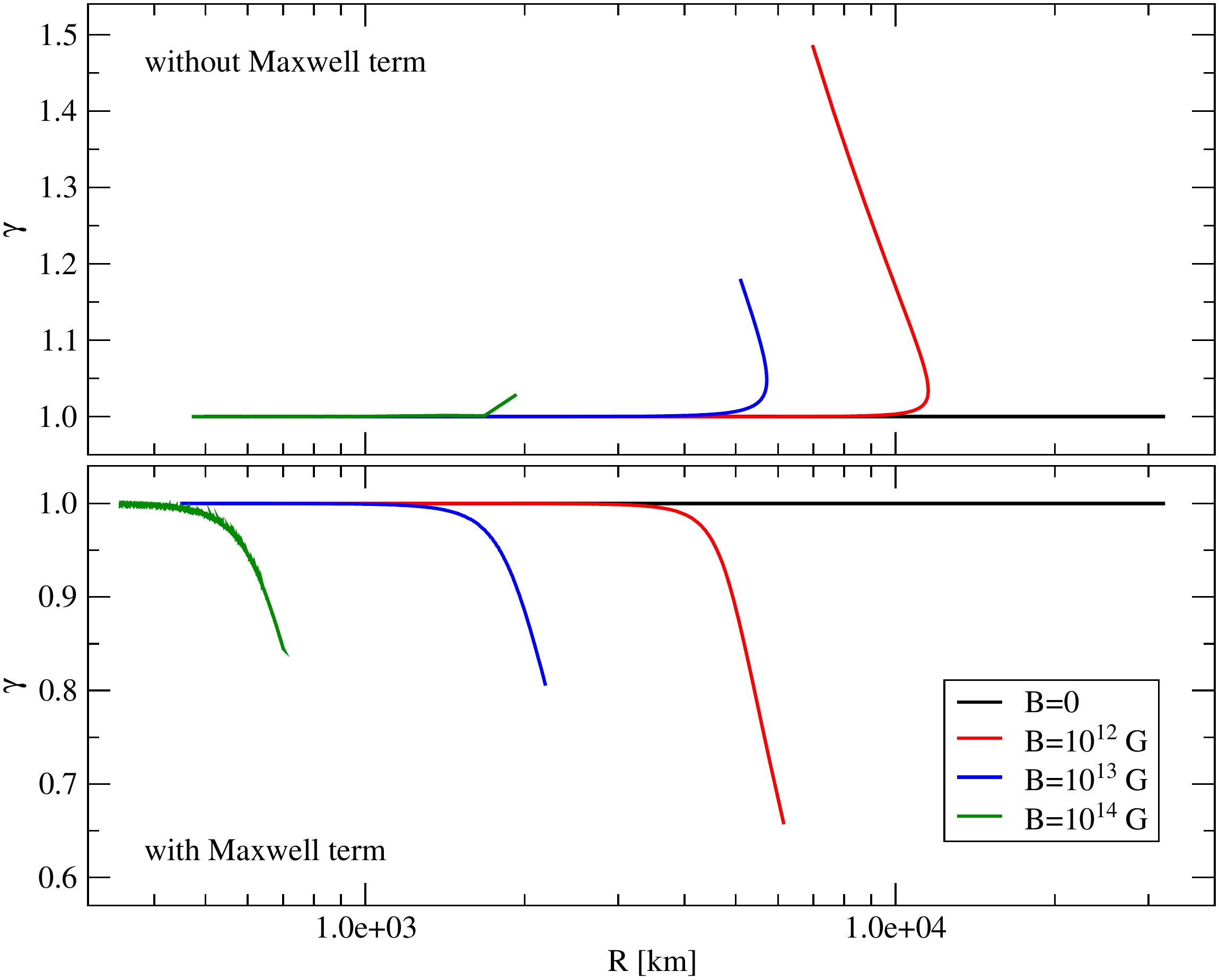}
	\caption{$\gamma$ parameter versus the equatorial radii. In the upper panel Maxwell term is neglected while in the lower panel it is considered.}
	\label{FigGR}
\end{figure}

\begin{figure}[ht]
	\includegraphics[width=0.48\textwidth]{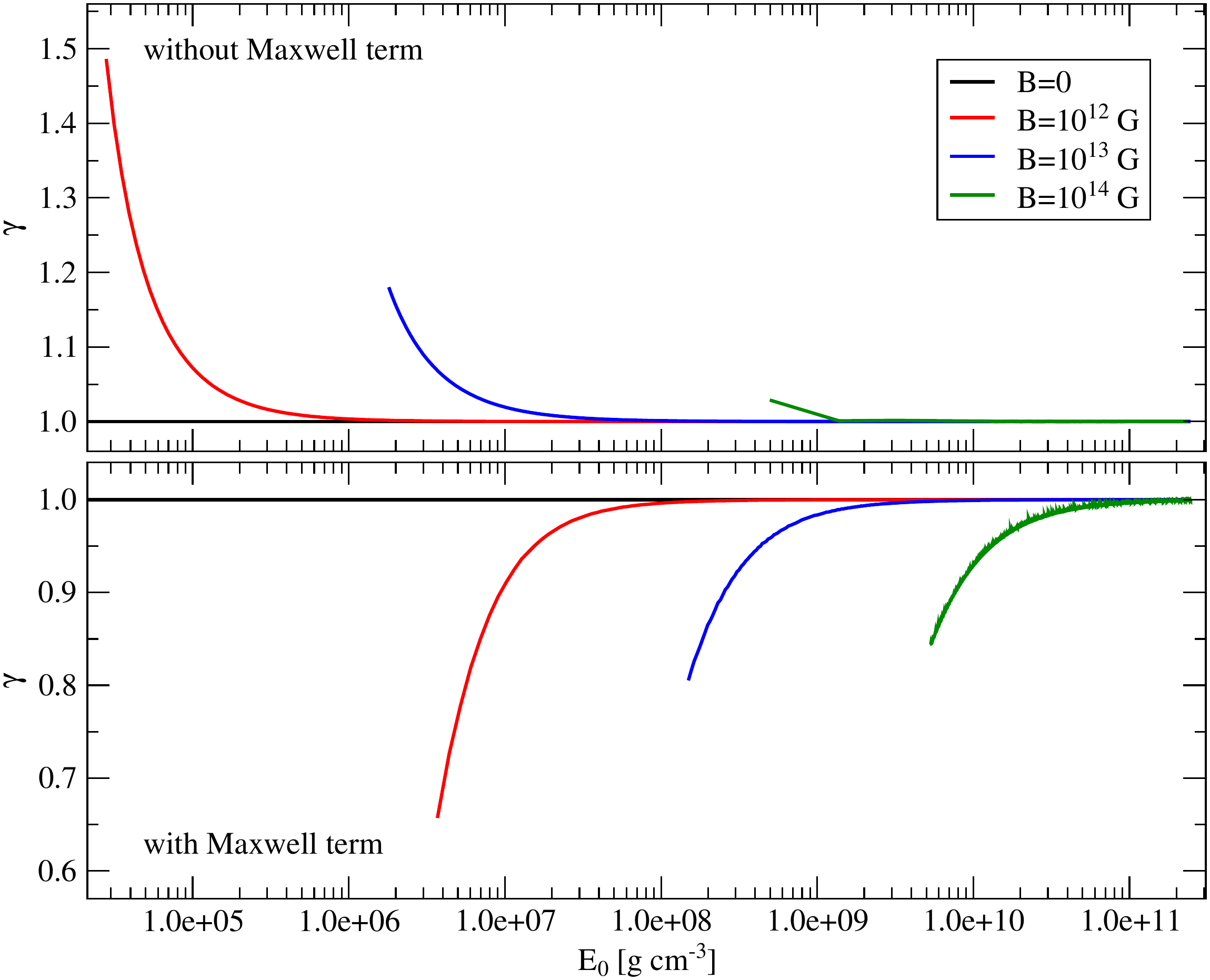}
	\caption{$\gamma$ parameter as a function of central energy density. In the upper panel Maxwell term is neglected while in the lower panel it is considered.}
	\label{FigGE0}
\end{figure}

In this section we validate the anisotropic model proposed in Section~\ref{sec3} by integrating  Eqs.~(\ref{gTOV}) for the EoS of magnetized WDs. The numerical results are shown in Figs.~\ref{FigMR},~\ref{FigME0},~\ref{FigGR} and \ref{FigGE0}.

Fig.~\ref{FigMR} displays the mass versus the equatorial $(R)$  ---the transverse and parallel ones--- for
$B=10^{12}~$G, $B=10^{13}~$G and $B=10^{14}~$G compared to the non-magnetized solution. We have considered the solutions with and without the Maxwell contribution to the pressures and the energy density.  For all values of the magnetic field at the highest central densities and smallest radii, the masses reach values close to the Chandrasekhar limit of $1.44~M_\odot$ \cite{chandrasekhar,shapiro,camenzind}. Also, note that in the $B=0$ case, the relation $R=Z$ is fulfilled and the curve is identical to the corresponding one in Fig.~\ref{FigIsotropicTOV}, as it should be, since Eqs.~(\ref{gTOV}) reduce to the isotropic TOV equations.

Moreover, without (with) the Maxwell contribution, an analysis of the solution at biggest radii, which corresponds to the lowest central densities, lead us to obtain a certain value of mass where the polar radius is higher (lower) than the equatorial one. So, the corresponding star is a prolate (oblate) object, as it is portrayed in the variation of $\gamma$ as a function of the central density $E_0$ (Fig.~\ref{FigGE0}), where the limiting values are shown for each case in Table \ref{tab:data}. This result also illustrates the the relation among the ansatz in Eq.~(\ref{gamma}) with the central density and the radius of the stars.

The existence of two values of mass for a given equatorial radius in the upper panel of Fig.~\ref{FigMR} can be understood as a deformation effect for low enough densities with respect to the magnetic energy of the system. If comparing to the corresponding curve of mass as a function of energy density in the lower panel on Fig.~\ref{FigME0}, we can see that the two masses come from stars with different central densities, so that the lower density star corresponds to the lower mass.

In the latter case, the magnetic field plays an important role, producing a higher deformation on the star. This can also be explained by the balance of the forces at stake, the magnetic, the gravitational and the one from the pressure exerted by the electron gas. For a given magnetic field, at the lowest densities, the particles can be more easily arranged in the direction of ${\bf B}$, so that the star is more deformed than another one with higher density and mass.

Therefore, the magnetic field effects becomes relevant at the low and intermediate energy density regime with respect to the value of the magnetic field considered and can be practically neglected for high densities. In consequence, the main effect is the deformation of the magnetized low density WDs. Relating this result with Fig.~\ref{FigIsotropicTOV}, we realize that the deformation could be seen on the TOV solution, but the loss of information due to the isotropic approximation was preventing any further conclusion on this matter.

\subsection{Stability and super-Chandrasekhar masses}

An important remark about the previous solutions is the fact that once again, we do not obtain masses above the Chandrasekhar limit \cite{1674-4527-15-10-1735}.

\begin{figure}[!ht]
	\includegraphics[width=0.48\textwidth]{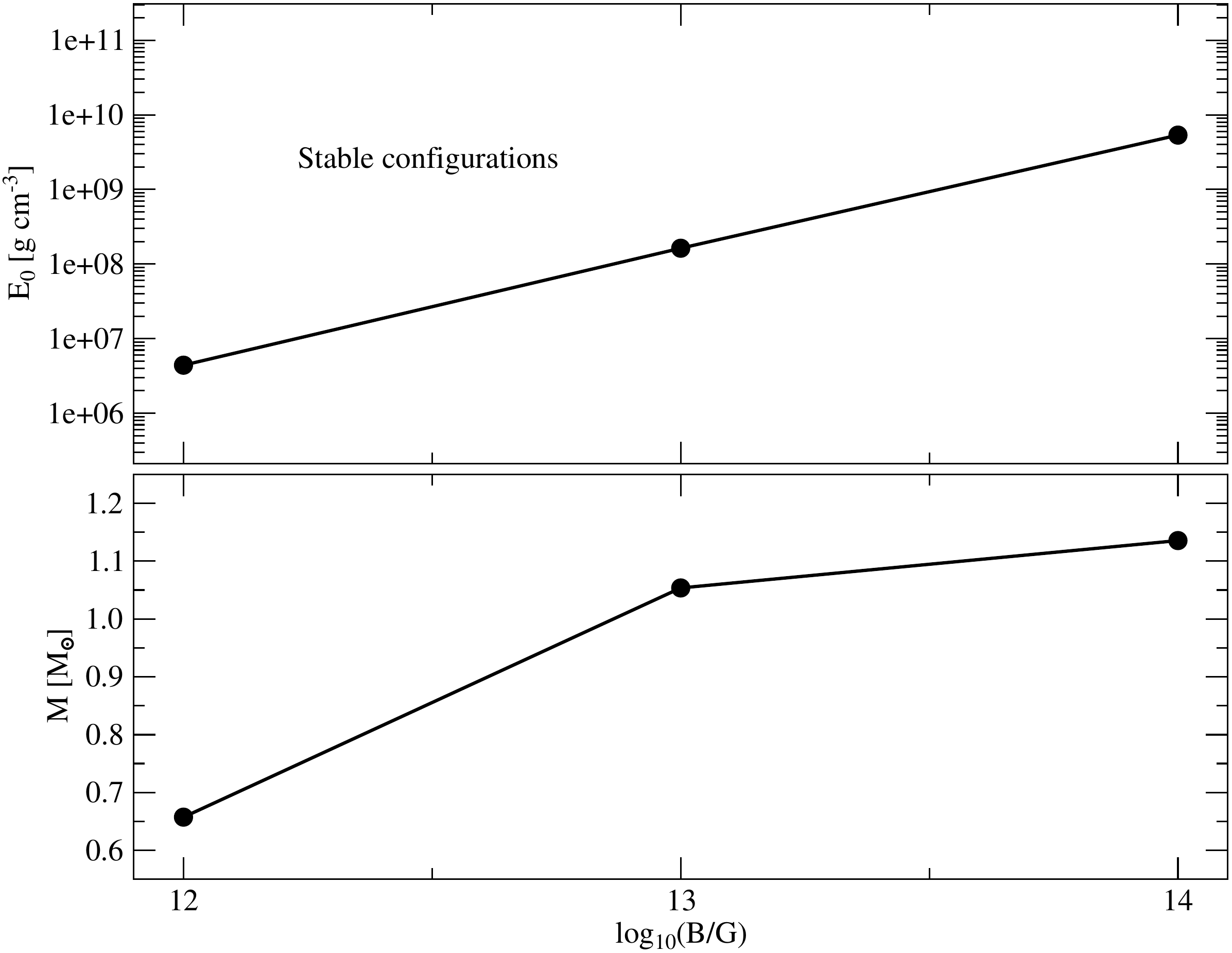}
	\caption{Onset of instability as a function of B. Central energy density (upper panel) and corresponding mass (lower panel) of the configuration below which $dM/dE_0<0$  for the solutions with Maxwell term.}
	\label{FigOnset}
\end{figure}

For the solutions obtained with the Maxwell term, at each value of the magnetic field, there is a minimum mass below whose corresponding central energy density the relation $dM/dE_0 <0$ is satisfied (lower panel of Fig.~\ref{FigME0}). This defines an onset for the central energy density below which solutions are unstable (upper panel of Fig.~\ref{FigOnset}). Table~\ref{tab:data} shows the corresponding data. Note that as the magnetic field increases, the onset density and mass become higher.

In the case without the Maxwell contribution, however, we can not make this analysis, since the mass versus central energy density curves do not display a region where $dM/dE_0 <0$ (upper panel of Fig.~\ref{FigME0}) for the densities relevant to this work. Nevertheless, the effect of fewer central densities that can account for stable configurations as $B$ increases remain, because the compromise of $\gamma$ parameter close to 1  must be respected. In this regard, note in Table~\ref{tab:data} that for $10^{12}~$G without Maxwell contribution there is almost no deformation  ($\gamma =1.0033$) at $E_0 \sim 10^6~$g~cm$^{-3}$ while for $\sim 10^{4}~$g~cm$^{-3}$ we get $\gamma=1.48$, which is an unphysical solution below the energy density range of interest (see corresponding curves in Figs.~\ref{FigMR},~\ref{FigME0},~\ref{FigGR} and \ref{FigGE0}).

One conclusion that comes out of such analysis is that if we move towards higher values of the magnetic field, of the order of $10^{15}- 10^{18}$ G, which are precisely the values employed in the works obtaining super-Chandrasekhar masses (Ref.~\cite{Das:2013gd} for instance), not only the effect of the magnetic field is magnified, but also the densities corresponding to stable objects go above the WDs density range.
\begin{widetext}
\begin{center}
\begin{table}[ht]
\caption{Values of $\gamma$, central energy density, equatorial and polar radii and mass for the most deformed configurations, which determine the onset of instability at $10^{12}~$G, $10^{13}~$G and $10^{14}~$G, in all cases ignoring and considering Maxwell term.}
	\begin{tabular}{cclccccc}\toprule
		\multicolumn{2}{c}{$B$ [G]} & Maxwell term& $\gamma$ &$E_0$ [g~cm$^{-3}$]& $R$ [km]& $Z$ [km]&$M$ [$M_{\odot}$] \\ \hline
		\multicolumn{2}{c}{\multirow{3}{*}{$10^{12}$}} &without&
		$1.0033$&$1.03519\times10^6$&$9988.1$&$10021.0$&$0.379$ \\ \cline{4-8}
		\multicolumn{2}{c}{\multirow{2}{*}{}} &&
		$1.4864$&$2.79605\times10^4$&$6973.6$&$10365.3$&$0.017$ \\ \cline{3-8}
		\multicolumn{2}{c}{}&with& $0.7267$&$4.40397\times10^6$&$5778.9$&$4199.4$&$0.657$ \\ \hline
		\multicolumn{2}{c}{\multirow{2}{*}{$10^{13}$}} &without&  $1.1802$&$1.80308\times10^6$&$5096.4$&$6014.7$&$0.248$\\  \cline{3-8}
		\multicolumn{2}{c}{}&with& $0.8259$&$1.62578\times10^8$&$2141.1$&$1768.3$&$1.054$ \\ \hline
		\multicolumn{2}{c}{\multirow{2}{*}{$10^{14}$}} &without&  $1.0289$&$4.97109\times10^8$&$1928.5$&$1984.2$&$1.122$\\  \cline{3-8}
		\multicolumn{2}{c}{}&with& $0.8458$&$5.34925\times10^9$&$699.8$&$591.9$&$1.135$ \\  \botrule
	\end{tabular}
\label{tab:data}
\end{table}
\end{center}
\end{widetext}

\section{Conclusions}\label{sec6}

In this work, we have obtained  the structure of non isotropic compact objects starting from a $\gamma$-metric and computing the mass as for a spheroid. As a result, we get a set of equations that describe the structure of an axially symmetric deformed object, provided it is spheroidal.
	
In the process to obtain the structure equations, we have neglected the dependence of the quantities with the angular coordinates. This means that when integrating the equations there is a lack of information and the total mass must be computed averaging the energy densities in the polar and equatorial direction. Then, a complete description of an axially symmetric object would require to consider dependence with all coordinates, which brings the necessity of more sophisticated numerical relativity techniques. However, the structure equations we present have the advantage of providing relevant information about the axially symmetric system at a low computational cost.

As we were interested in the anisotropies coming from magnetic fields effects on compact objects, we have connected the parameter $\gamma$, which relates the radii of the spheroid, with the source of the anisotropy through the ratio between the central pressures, thus linking the physics determining the properties of the matter that composes the star to its shape. For the validity of the ansatz, $\gamma$ parameter must be close to 1 to produce slight modifications in the energy densities and obtain physical results.

In order to illustrate our model we solved the modified structure equations to obtain magnetized WDs structure considering magnetic field values of $10^{12}$~G, $10^{13}$~G and $10^{14}$~G and densities from $10^6-10^{11}~\text{g}/\text{cm}^3$. Solutions were presented ignoring and considering the Maxwell term in pressures and energy densities.  If this contribution is included, it wins over the matter term and inverts the behavior of perpendicular and parallel pressures. This choice allowed us to have two sets of EoS, one where $\gamma>1$, and the other one with $\gamma<1$.

Due to the constant character of the Maxwell contribution, the net effect it produces when considered in the pressures and energy density are: to change the form of the deformation of stable configurations from prolate to oblate; to increase the deformation and to shift towards higher energy densities the region when the magnetic field effects become relevant.

Our results show that the effect of the magnetic anisotropy on the EoS is relevant at the low and intermediate density regime with respect to the magnetic field  in both cases. Besides, the magnetic field does not affect maximum values of WDs' masses. The observed effect is the prolate/oblate deformation of stable magnetized WDs configurations with respect to the corresponding central densities solutions in absence of magnetic field.

As  $\gamma$-structure equations are general, they can be useful to study other types of magnetized compact objects.


\section*{Acknowledgments}

The author thanks to anonymous referees for their valuable comments which enrich the manuscript. D.A.T, S.L.P, D.M.P, A.P.M and G.Q.A have been supported by the grant No. 500.03401 and V.H.M by the grant No. 500.03501, both of PNCB-MES, Cuba. D.M.P  has been also supported by a DGAPA-UNAM fellowship.

\end{document}